\documentclass[lettersize,journal]{IEEEtran}
\usepackage{amsmath,amsfonts}
\usepackage{algorithmic}
\usepackage{algorithm}
\usepackage{array}
\usepackage[caption=false,font=normalsize,labelfont=sf,textfont=sf]{subfig}
\usepackage{textcomp}
\usepackage{stfloats}
\usepackage{url}
\usepackage{verbatim}
\usepackage{graphicx}
\usepackage{cite}
\usepackage{hhline}
\usepackage{enumitem}
\usepackage{multirow}
\usepackage[table]{xcolor}
\usepackage{rotating}
\usepackage{makecell}
\newcommand{\myrotcell}[1]{\rotcell{\makebox[0pt][l]{#1}}}

\usepackage{url}
\usepackage{accents}
\usepackage{tensor}

\usepackage{mathtools,cuted}
\usepackage{amsmath,amsfonts}
\usepackage{algorithmic}
\usepackage{algorithm}
\usepackage{array}
\usepackage[nolist]{acronym}
\usepackage{etoolbox}
\usepackage[pdfpagelabels,  
bookmarks,       
pdftex
]{hyperref}
\hypersetup{
	pdfborder   = 0 0 0,
	plainpages  = false,
	bookmarksnumbered = true,
}
\usepackage{color}
\PassOptionsToPackage{x11names}{xcolor}
\usepackage{xcolor}
\usepackage{orcidlink}
\usepackage{framed}
\usepackage{flushend}

\newcommand{\bi}[1]{\boldsymbol{#1}}
\newcommand{\ur}[1]{\mathrm{#1}}
\newcommand{\cali}[1]{\mathcal{#1}}

\newcommand\norm[1]{\left\lVert#1\right\rVert}

\newcommand*{\tr}{^{\top}}

\newcommand*{\matlab}{\textsc{Matlab}\,}

\newcommand*{\intelcore}{\textsc{Intel}\textsuperscript{\copyright} \textsc{Core}\textsuperscript{TM} $\:$}
\newcommand*{\simulink}{\textsc{Simulink}\,}
\newcommand*{\simpack}{\textsc{Simpack}\,}

\newcommand*{\casadi}{\textsc{CasADi}\,}
\mathchardef\mhyphen="2D

\newcommand\CF[1]{ { \ur{(CF)}_{ \ur{#1} } } }

\newcommand{\eg}{e.\,g.,\,}
\newcommand{\ie}{i.\,e.,\,}
\newcommand*{\lagrange}{\textsc{Lagrange}\,}
\newcommand*{\euler}{\textsc{Euler}\,}
\newcommand*{\lyapunov}{\textsc{Lyapunov}\,}
\newcommand{\psitrax}{\psi_{\ur{TrAx}}}
\newcommand{\ytrax}{y_{\ur{TrAx}}}
\newcommand{\ps}[3][]{\bar{\bi{#2}}_{\left[#3\right]}^{\mathrm{#1}}(t)} % predictive sequence

\def\relPath{figures/} % initialize path
\def\extendmyPath#1#2{\edef\fullPath{#1\relPath#2}} % command for inserting figure name

\def\figLabel{Fig.} % initialize 
\def\extendmyfigLabel#1#2{\edef\fullfigLabel{#1\figLabel#2}} % command for inserting figure name

\def\tabLabel{Tab.} % initialize 
\def\extendmytabLabel#1#2{\edef\fulltabLabel{#1\tabLabel#2}} % command for inserting figure name

\newcommand{\newfig}[4][]{
	\extendmyPath{}{#3}
	\extendmyfigLabel{}{#3}
	\begin{figure}[#1]
		\centering %
			\includegraphics[#2]{\fullPath}
			\caption{#4} \label{\fullfigLabel}
		\end{figure}
	}
	
\newcommand{\newtab}[5][]{
	\extendmytabLabel{}{#4}
	\begin{table}[#1]
		\centering
		\caption{#5}
		\label{\fulltabLabel}
		\def\arraystretch{1.2}%
		\begin{tabular}{#2}
			#3
		\end{tabular}
	\end{table}
}
	
\newcommand{\fig}[1]{\extendmyfigLabel{}{#1}Fig.~\ref{\fullfigLabel}}    % Zitieren von Bildern
\newcommand{\eqn}[1]{(\ref{#1})}   % Zitieren von Gleichungen
\newcommand{\tab}[1]{\extendmytabLabel{}{#1}Tab.~\ref{\fulltabLabel}}  % Zitieren von Tabellen

\newcommand{\secref}[1]{Section~\ref{#1}}

\newcommand\copyrighttext{%
	\footnotesize \copyright 2024 IEEE.  Personal use of this material is permitted. Permission from IEEE must be obtained for all other uses, in any current or future media, including reprinting/republishing this material for advertising or promotional purposes, creating new collective works, for resale or redistribution to servers or lists, or reuse of any copyrighted component of this work in other works.}
\newcommand\copyrightnotice{%
	\begin{tikzpicture}[remember picture,overlay]
		\node[anchor=south,yshift=10pt] at (current page.south) {\fbox{\parbox{\dimexpr\textwidth-\fboxsep-\fboxrule\relax}{\copyrighttext}}};
	\end{tikzpicture}%
}

\begin{acronym}
	\acro{rnn}[RNN]{Recurrent Neural Network}
	\acroplural{rnn}[RNN]{Recurrent Neural Networks}
	\acro{narx}[NARX-NN]{Nonlinear Autoregressive Exogenous Neural Network}
	\acroplural{narx}[NARX-NN]{Nonlinear Autoregressive Exogenous Neural Networks}
	\acro{gru}[GRU]{Gated Recurrent Unit}
	\acroplural{gru}[GRU]{Gated Recurrent Units}
	\acro{lstm}[LSTM]{Long Short-Term Memory}
	\acro{ann}[ANN]{Artificial Neural Network}	
	\acroplural{ann}[ANN]{Artificial Neural Networks}
	\acro{ffnn}[FFNN]{Feedforward Neural Network}
	\acroplural{ffnn}[FFNN]{Feedforward Neural Networks}
	
	\acro{kf}[KF]{Kalman Filter}	
	\acroplural{kf}[KF]{Kalman Filters}
	\acro{ekf}[EKF]{Extended Kalman Filter}
	\acro{nekf}[NEKF]{Neural Extended Kalman Filter}
	\acro{ukf}[UKF]{Unscented Kalman Filter}
	\acro{knn}[$K$NN]{$K$-Nearest-Neighbors}
	\acro{mftm}[MFTM]{Magic Formula Tire Model}
	\acro{pso}[PSO]{Particle Swarm Optimization}
	\acro{sqp}[SQP]{Sequentielle Quadratische Programmierung}
	
	\acro{cog}[COG]{Center Of Gravity}
	\acro{ebs}[EBS]{Electronic Braking System}
	\acro{lti}[LTI]{Linear Time-Invariant}
	\acro{pso}[PSO]{Particle Swarm Optimization}
	\acro{nmse}[NMSE]{Normalized Mean Squared Error}
	\acro{mse}[MSE]{Mean Squared Error}
	\acro{rmse}[RMSE]{Root Mean Squared Error}
	
	\acro{rpe}[RPE]{Recursive Predictive Error}
	\acro{adas}[ADAS]{Advanced Driver Assistance Systems}
	
	\acro{ngt}[NGT]{Next Generation Train}
	\acro{dlr}[DLR]{German Aerospace Center}
	
	\acro{mpc}[MPC]{Model Predictive Control}
	\acro{nmpc}[NMPC]{Nonlinear Model Predictive Control}
	\acro{ltv}[LTV]{Linear Time-Variant}
	\acro{ltvmpc}[LTV-MPC]{\acl{ltv} Model Predictive Control}
	\acro{ndi}[NDI]{Nonlinear Dynamic Inversion}
	
	\acro{irw}[IRW]{Independently Rotating Wheels}
	\acro{dirw}[DIRW]{Driven \acl{irw}}
	\acro{mbs}[MBS]{Multi-Body Simulation}
	\acroplural{mbs}[MBS]{Multi-Body Simulations}
	
	\acro{sr}[SR]{Institut für Systemdynamik und Regelungstechnik}
	\acro{db}[DB]{Deutsche Bahn}
	\acro{ice}[ICE]{Intercity-Express}
	
	\acro{cf}[CF]{Coordinate Frame}	
	\acroplural{cf}[CF]{Coordinate Frames}
	
	\acro{gnss}[GNSS]{Global Navigation Satellite System}
	\acroplural{gnss}[GNSS]{Global Navigation Satellite Systems}
\end{acronym}

	\newcommand{\mytitle}{Integrated Model Predictive Control of High-Speed Railway Running Gears with Driven Independently Rotating Wheels}
	\newcommand{\mykeywords}{Model Predictive Control, integrated control, adhesion control, railway vehicle dynamics, independently rotating wheels}
	\newcommand{\myauthors}{Jan-Hendrik~Ewering\,\orcidlink{0000-0002-4095-9705}, Christoph~Schwarz\,\orcidlink{0000-0003-4882-0828}, Simon~F.~G.~Ehlers\,\orcidlink{0000-0001-5524-6639}, Hans-Georg~Jacob, Thomas~Seel\,\orcidlink{0000-0002-6920-1690}, and Andreas~Heckmann\,\orcidlink{0000-0003-3883-8653}}
	
	\hypersetup{
		pdftitle    = {\mytitle},
		pdfsubject  = {\mytitle}, 
		pdfauthor   = {\myauthors},
		pdfkeywords = {\mykeywords} 
	}
\begin{document}
		\title{\mytitle}
		
		\author{\myauthors
			% <-this % stops a space
			\thanks{Jan-Hendrik Ewering, Simon Ehlers, Hans-Georg Jacob, and Thomas Seel are with the Institute of Mechatronic Systems, Leibniz Universit\"at Hannover, 30167~Hanover, Germany (e-mail: \href{mailto:jan-hendrik.ewering@imes.uni-hannover.de}{jan-hendrik.ewering@imes.uni-hannover.de}).}%
			\thanks{Christoph Schwarz, and Andreas Heckmann are with the Institute of System Dynamics and Control, German Aerospace Center (DLR), 82234~Wessling, Germany.}}
%			\thanks{Digital Object Identifier []}}
		
		\IEEEpubid{~}

		% Remember, if you use this you must call \IEEEpubidadjcol in the second column for its text to clear the IEEEpubid mark.

		\maketitle
		
		\begin{abstract}
			Railway running gears with \ac{irw} can significantly improve wear figures, comfort, and safety in railway transportation, but certain measures for wheelset stabilization are required. This is one reason why the application of traditional wheelsets is still common practice in industry. Apart from lateral guidance, the longitudinal control is of crucial importance for railway safety.\\
			In the current contribution, an integrated controller for joined lateral and longitudinal control of a high-speed railway running gear with driven \ac{irw} is designed. To this end, a novel adhesion-based traction control law is combined with \ac{ltv} and nonlinear \ac{mpc} schemes for lateral guidance. The \ac{mpc} schemes are able to use tabulated track geometry data and preview information about set points to minimize the lateral displacement error.\\ 
			Co-simulation results with a detailed \ac{mbs} show the effectiveness of the approach compared with state-of-the-art techniques in various scenarios, including curving, varying velocities up to 400 km/h and abruptly changing wheel-rail adhesion conditions.
		\end{abstract}
		
		
		\copyrightnotice
		
		% main text ============================================================
		\acresetall
		\section{Introduction}\label{sec:introduction}
		\IEEEPARstart{T}{he} climate crisis is one of the most urgent challenges of today and insistently demonstrates the necessity for a transition of the mobility sector. Increasing material prices, dependency on international actors, and rising transport volumes exacerbate the problem.\\
		In this context, railway transport can make a contribution to a sustainable and environmentally friendly mobility, and hence its further improvement is of major interest. A promising design option for railway running gears is to employ \ac{irw} instead of commonly used wheelsets. \ac{irw} introduce an additional degree of freedom to the system and allow (if actively controlled) to specify the exact lateral position of the running gear in the track. Thereby, wear figures and ride comfort can be improved dramatically \cite{Abdelfattah.2014}. In particular, undesired slip in the wheel-rail contacts, and hence wear, can be reduced by nearly ideal rolling. Further, the omission of the middle axle allows for new train designs such as low-floor or double-deck trains with continuous floors on both levels, which has implications for effectiveness and accessibility.\\
		In this context, the \ac{dlr} conducts railway research as part of the project \ac{ngt}, which aims towards the development of a future train concept. In detail, the train is equipped with distributed drives in each running gear which are holistically controlled. At the lowest level, each running gear with driven \ac{irw} should robustly perform mechatronic guidance and slip prevention independently of higher control levels.\\ 
		Multiple control strategies have been devised for mechatronic lateral guidance of running gears with \ac{irw}, some of which are based on a cascaded PID control structure \cite{Liu.2021}. Besides, the application of $H_{\infty}$ control ensured robust stabilization and guidance in the presence of parameter variations \cite{Mei.2001}. Other studies employed gain scheduled state space controllers due to the strongly velocity-dependent lateral dynamics \cite{Mei.2003}. Additional physically motivated feedforward control signals can further improve the performance, and a parameter space approach has been used to ensure robustness despite the nonlinear wheel and rail profiles \cite{AndreasHeckmann.2016,AndreasHeckmann.2017}. Since running gears with \ac{irw} show strongly nonlinear behavior, other advanced control strategies, such as feedback linearization, are suitable choices as well \cite{Grether.2018}.\\ \IEEEpubidadjcol
		In longitudinal control, recent research has focused on the reduction of braking distances. In this context, some groups proposed slip-based longitudinal control. These methods depend on a desired slip area in which the best braking performance (\ie the maximum adhesion) is assumed to be \cite{Lee.2016,Stutzle.2008}. However, adhesion conditions between wheels and rails are highly variable and depend on environmental influences. Hence, the maximum of the adhesion-slip characteristic is located at changing slip values. A possibility to overcome this problem is the application of maximum-seeking controllers. An example, which is based on a sliding mode approach, is shown in \cite{Schwarz.2021,Schwarz.2022}. Crucial for this concept is the availability of reliable real-time slip and adhesion estimates. In the given example, the estimates were obtained from an extended Kalman filter \cite{Schwarz.2020}.\\
		In the context of combined lateral and longitudinal control, an integration approach is required for two main reasons: First, lateral and longitudinal system dynamics are inherently coupled in traditional vehicles with nonholonomic properties. Second, lateral guidance and longitudinal traction or braking forces are all adhesion-dependent and occur in the contacts between wheels and underground. Therefore, they are limited by the current adhesion conditions. If the combined lateral and longitudinal adhesion exceeds the maximum possible adhesion in a contact, extensive slip and hence instability occurs (\ie locking or skidding of the wheels).\\
		For automotive systems, many examples of successful implementations of such integrated longitudinal and lateral controllers can be found in the literature. A popular methodological choice is \ac{mpc} where an optimization problem is solved repetitively to determine the control input based on a predictive plant model \cite{Falcone.2007,Ataei.2020,Zhao.2015}.\\
		Regarding the integrated control of railway running gears with \ac{irw}, only first attempts have been made so far. In \cite{Gretzschel.2002} and \cite{Perez.2004}, a differential torque determined by a lateral guidance controller was added to a superimposed longitudinal traction or braking torque. This integration approach can be considered a decentralized overall control system \cite{Gordon.2003}. A similar integration strategy with more sophisticated subsystem controllers was used in other contributions \cite{Feng.2008,Lu.2017}.\\
		All presented methods suffer from the fact that a fixed allocation of torques is needed for both subsystems in order to comply with the actuator saturation. As an example, the lateral controller does not require the complete allocated torque while riding on a straight track with few irregularities. If the emergency brake is applied in this scenario, only the amount of torque allocated to the longitudinal controller can be used for braking, even if more motor torque could be provided.\\
		A possible solution is the introduction of a supervised integration strategy in the form of a variable allocation rule. This technique is proposed in \cite{Schwarz.2022b} and can reduce the braking distance in the presented simulation scenario by almost $5\%$. However, a custom and smooth shift between emphasis on lateral or longitudinal control is not possible in \cite{Schwarz.2022b}.\\
		To the best of the authors' knowledge, \ac{mpc} has not been used for control of railway running gears with \ac{irw} so far. However, its application is appealing since,
		\begin{itemize}[noitemsep,labelindent=\parindent,leftmargin=20pt,topsep=0pt]
			\item nonlinear and coupled system dynamics can be employed,
			\item preview information about track geometry and set-points can be considered in the prediction horizon,
			\item individual cost functions and weightings can be used to adjust the control objective for different scenarios.
		\end{itemize}
		
		\noindent In this light, we contribute an integrated controller featuring two subsystems, an \ac{mpc} scheme for lateral control and a novel adhesion controller for longitudinal control. 
		Thus, a holistic control framework for joined longitudinal and lateral control of railway running gears with \ac{irw} is introduced and evaluated. 
		The contribution of this study is threefold:
		\begin{enumerate}[noitemsep,labelindent=\parindent,leftmargin=20pt,topsep=0pt]
			\item For the first time, \ac{mpc} is employed for lateral guidance of railway running gears with \ac{irw}. Beyond the current state of research, the predictive nature of \ac{mpc} is exploited by considering known track geometry and known set point trajectories in the running gear controller explicitly.
			\item A novel longitudinal control law is derived. It builds upon an existing maximum-seeking controller for poor adhesion conditions and is extended for autonomous use in general operation without knowledge of the adhesion conditions between wheels and rails.
			\item Few integrated controllers for railway running gears with \ac{irw} exist. Beyond the current state of research, the proposed integrated controller allows for a custom shift between an emphasis on lateral or longitudinal control by leveraging the flexibility of \ac{mpc}.
		\end{enumerate}
		
		\noindent The conceptual advantages of the proposed integrated control scheme are illustrated and quantitatively described by means of a detailed \ac{mbs}. Both, the lateral and longitudinal control accuracy are improved compared to a state-of-the-art controller for railway running gears with \ac{irw}.\\
		In \secref{sec:modeling}, the running gear system is introduced and modeled with regard to the control context. The control synthesis is outlined in \secref{sec:controller-synthesis} before experimental results of the \ac{mbs} are presented in \secref{sec:experimental-results}. The control scheme of \cite{Grether.2018,Heckmann.2020}, which is based on feedback linearization, is given as a comparison. Last, we discuss our findings in \secref{sec:discussion} and give a conclusion together with an outlook for future research (\secref{sec:conclusion}).
		
		\section{Modeling}\label{sec:modeling}
		To start with, the running gear system is introduced. The dynamics of a running gear system with \ac{irw} have been analyzed thoroughly in the past, see \cite{Mei.2001,Goodall.2000}. Only the main influences on the system are described for brevity and to provide a background for the subsequent control model. Relevant variables and frames are described, and the modeling of the track is explained in order to create a framework for consideration of preview information about the track path over the prediction horizon. For use in \ac{mpc}, a simplified system model is derived. 
		
		\subsection{Coordinate Frames}\label{sec:system-description}
		A system overview with relevant \acp{cf} can be found in \fig{figure_1}. Here, the world \ac{cf} has a fixed location and orientation and is denoted by subscript $\square_\ur{0}$.\\
		A track \ac{cf} $\CF{Tr}$ is defined to describe the position and orientation of the track. It locates at the current position $p$ along the track path, which is associated with a position vector in Cartesian coordinates $\bi{r}_{\ur{0Tr}}\tr = \left[	x_{\ur{Tr}} , y_{\ur{Tr}}, z_{\ur{Tr}} \right]$. The $x$-axis of $\CF{Tr}$ is aligned with the track center line, and its $y$-axis lies in the track plane (\ie the plane between left and right rail tops). This plane is not necessarily identical with the $x$-$y$-plane of $\CF{0}$ due to a rail superelevation of angle $\varphi_{\ur{Tr}}$ or a rail inclination of angle $\epsilon_{\ur{Tr}}$. The $z$-axis of $\CF{Tr}$ is orthogonal to the track plane and points downwards. The orientation of the track frame can be transformed into world coordinates by the rotation matrix
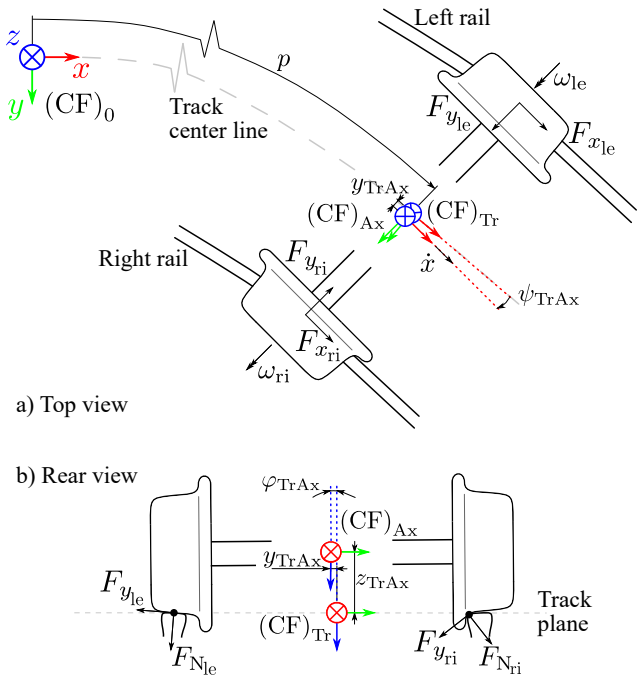
		\newfig[b]{width=8.4cm}{figure_1}{System overview of a railway running gear with \ac{irw}.}
		\begin{align}\label{eqn.rot_matrix}
			\bi{R}_{\ur{0Tr}}  = &\begin{bmatrix}
				\cos \psi_{\ur{Tr}} 	& -\sin \psi_{\ur{Tr}} 		& 0			\\
				\sin \psi_{\ur{Tr}}     & \cos \psi_{\ur{Tr}} 		& 0	\\
				0      					& 0 						& 1
			\end{bmatrix}
			\begin{bmatrix}
				\cos \epsilon_{\ur{Tr}} 	& 0		& \sin \epsilon_{\ur{Tr}}			\\
				0      					& 1 	& 0 \\
				-\sin \epsilon_{\ur{Tr}}     & 0		& \cos \epsilon_{\ur{Tr}}	
			\end{bmatrix} \nonumber\\
			&\begin{bmatrix}
				1 & 0 						& 0 \\
				0 & \cos \varphi_{\ur{Tr}} 	& -\sin \varphi_{\ur{Tr}}\\
				0 & \sin \varphi_{\ur{Tr}}     & \cos \varphi_{\ur{Tr}}
			\end{bmatrix},
		\end{align}

		\noindent where $\varphi_{\ur{Tr}}$ is the roll angle about the $x$-axis of $\CF{Tr}$,
		$\epsilon_{\ur{Tr}}$ is the pitch angle of the track (\ie track inclination), and  
		$\psi_{\ur{Tr}}$ is the yaw angle between $\CF{Tr}$ and $\CF{0}$, respectively.\\
		A third frame $\CF{Ax}$ is introduced to describe position and orientation of the axle. It is located at track coordinate $p$ as well, but its origin lies in the middle between the two wheel centers, as can be seen in the rear view in \fig{figure_1}. 
		Thus, the position offset between $\CF{Ax}$ and $\CF{Tr}$ can be described by $\bi{r}_{\ur{TrAx}}\tr = \left[	0 , \ytrax, z_{\ur{TrAx}} \right]$. In addition, the orientation of axle body frame and track frame can differ by a relative yaw angle $\psi_{\ur{TrAx}}$ and a relative roll angle $\varphi_{\ur{TrAx}}$. The corresponding rotation matrix between $\CF{Ax}$ and $\CF{Tr}$ is
		\begin{align}
			\bi{R}_{\ur{TrAx}} = &\begin{bmatrix}
				\cos \psi_{\ur{TrAx}} 	& -\sin \psi_{\ur{TrAx}} 		& 0			\\
				\sin \psi_{\ur{TrAx}}     & \cos \psi_{\ur{TrAx}} 		& 0	\\
				0      					& 0 						& 1
			\end{bmatrix} \nonumber\\
			&\begin{bmatrix}
				1 & 0 						& 0 \\
				0 & \cos \varphi_{\ur{TrAx}} 	& -\sin \varphi_{\ur{TrAx}}\\
				0 & \sin \varphi_{\ur{TrAx}}     & \cos \varphi_{\ur{TrAx}}
			\end{bmatrix}.
		\end{align}
		In this light, $\ytrax$ and $\psitrax$ are measures of the lateral deviation of the axle from the track center line. In contrast, the quantities $z_{\ur{TrAx}}$ and $\varphi_{\ur{TrAx}}$ describe the relative vertical and roll angle deviation, respectively. Assuming nonelastic materials, knowledge of the wheel and rail profiles and that the wheels do not lift from the rails, both values $z_{\ur{TrAx}}$ and $\varphi_{\ur{TrAx}}$ can be inferred from other variables, meaning they are entirely dependent and can be calculated deterministically.
		
		\subsection{Train-Track Interaction}
		For use of track preview data in \ac{mpc}, orientation angles of the track frame are tabulated dependent on the track coordinate $p$ and in accordance with \eqn{eqn.rot_matrix}, \ie
		\begin{subequations}\label{track}
			\begin{align}
				&\text{Absolute curve angle } &&{} \psi_{\ur{Tr}}(p), \\ 
				&\text{Absolute curve rate } &&{} \dot{\psi}_{\ur{Tr}}(p) = \left. \frac{\partial \psi_{\ur{Tr}}}{\partial p} \right|_{p} \frac{\ur{d} p}{\ur{d} t}, \label{tracka}\\ 
				&\text{Superelevation angle } &&{}  \varphi_{\ur{Tr}}(p), \\ 
				&\text{Superelevation rate } &&{}  \dot{\varphi}_{\ur{Tr}}(p) = \left. \frac{\partial \varphi_{\ur{Tr}}}{\partial p} \right|_{p} \frac{\ur{d} p}{\ur{d} t}, \label{trackb}\\ 
				&\text{Inclination angle } &&{}  \epsilon_{\ur{Tr}}(p),\\ 
				&\text{Inclination rate } &&{}  \dot{\epsilon}_{\ur{Tr}}(p) = \left. \frac{\partial \epsilon_{\ur{Tr}}}{\partial p} \right|_{p} \frac{\ur{d} p}{\ur{d} t}, \label{trackc}
			\end{align}
		\end{subequations}
		which is summarized by means of the expression $\cali{D}_{\ur{Tr}}~=~\begin{Bmatrix}
			\psi_{\ur{Tr}} & \frac{\partial \psi_{\ur{Tr}}}{\partial p} & \varphi_{\ur{Tr}} & \frac{\partial \varphi_{\ur{Tr}}}{\partial p}  &  \epsilon_{\ur{Tr}} & \frac{\partial \epsilon_{\ur{Tr}}}{\partial p}
		\end{Bmatrix} $. Please note, the velocity along the track path is needed for calculation of the current rotational velocities of the track frame. To this end, the time derivative $\frac{\ur{d} p}{\ur{d} t}$ is approximated by $\dot{x}$. Additional track irregularities occur superimposed to the ideal track path \eqn{track} and can be described by $p$-dependent functions as well.\\
		For description of the running gear system itself, the generalized coordinates $\dot{\bi{q}}\tr = \left[ \dot{x}, \dot{y},\dot{\psi}_{\ur{Ax}}, \omega_j \right]$ can be employed, which are defined in the axle body frame $\CF{Ax}$. The subscript $\square_j$, $j \in \left\{\ur{le}, \ur{ri}\right\}$, refers to the left and right side of the running gear, respectively.\\
		In addition, several external forces and torques are acting on the running gear. First, the motor torques $\tau_{j}$ are acting on the right and left wheel, respectively. Second, the train body causes forces $\cali{F}_{\ur{Train}}$ on the running gear. These are mitigated by a suspension stage in the form of dampers and springs placed between car body and wheel carrier. The corresponding spring and damper forces have a major impact on rolling stability of the running gear. Last, contact forces and moments occur in the contact points between wheels and rails. The contact forces in normal ($F_{\ur{N}}$), lateral ($F_{y}$) and longitudinal direction ($F_{x}$) of the corresponding left and right wheel-rail contact are visualized in \fig{figure_1}.\\
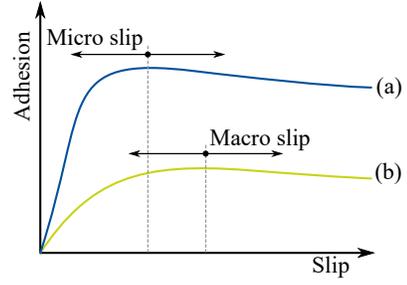
		\newfig[t]{width=5.2cm}{figure_2}{Schematic adhesion-slip characteristics with good (a) and poor (b) adhesion conditions.}
		A more detailed analysis of these contact forces is presented subsequently, since the lateral and longitudinal running behavior of the running gear is inherently coupled by the forces acting in the contact points. These forces are related to the current adhesion conditions, which is usually described in terms of the adhesion-slip characteristic in the contact points (see examples in \fig{figure_2}). The momentary adhesion-slip characteristic can be represented by means of different models that employ a set of suitable model parameters $\cali{C}_{\ur{Tr}}$. 
		Those models describe the longitudinal and lateral adhesion ($f_x$ and $f_y$) dependent on longitudinal slip, lateral slip and spin creepage ($s_x$, $s_y$ and $s_{\psi}$) according to some function
		\begin{equation}\label{eqn.kalker}
			\begin{bmatrix}
				f_x \\ f_y
			\end{bmatrix}
			= \bi{h}\left(
			\begin{bmatrix}
				s_x \\ s_y \\ s_{\psi}
			\end{bmatrix}, \cali{C}_{\ur{Tr}}\right).
		\end{equation}
		The absolute lateral and longitudinal contact forces are dependent on the normal contact force $F_{\ur{N}}$ and read $F_x = F_{\ur{N}} f_x$ and $F_y = F_{\ur{N}} f_y$, respectively. In \eqn{eqn.kalker}, the creep torque about the vertical axis is neglected.\\
		It is worth mentioning that the Euclidean sum of the slip in lateral and longitudinal direction is bounded. The same holds for the adhesion, \ie 
		\begin{align}
			s_{\ur{max}} &\geq s = \sqrt{s_x^2 + s_y^2},\\
			f_{\ur{max}} &\geq f = \sqrt{f_x^2 + f_y^2},
		\end{align}
		where $s$ is the total slip, and $f$ is the total tangential adhesion, respectively. Hence, all generalized coordinates are affected by contact forces and, in turn, adhesion.
		
		\subsection{Control-Oriented Model}\label{sec:control-oriented-model}
		For use in \ac{mpc}, a simplified and discrete-time system representation $\bi{f}_{\ur{d}}: \mathbb{R}^{n} \times \mathbb{R}^{m} \rightarrow \mathbb{R}^{n}$ with
		\begin{equation}\label{eq.discr_state_space1}
			\bi{x}_{k+1} = \bi{f}_{\ur{d}} \left(\bi{x}_k,\bi{u}_k, \theta \right),
		\end{equation}
		is derived by means of the \lagrange formalism. It features the state vector $\bi{x} \in \mathbb{R}^{n}$, the input vector $\bi{u} \in \mathbb{R}^{m}$, and a set which summarizes the varying parameters $\theta$. A schematic can be seen in \fig{figure_3}. It is worth mentioning that the control model should only capture major dynamics over a short time horizon.\\
		To account for the movement of the running gear along and inside the track, two additional variables $\ytrax$ and $\psitrax$ have been introduced, which describe relative displacement and yaw angle between $\CF{Tr}$ and $\CF{Ax}$, respectively. Therefore, $\ytrax$ and $\psitrax$ are measures for the deviation from the desired path and should be minimized by the lateral guidance controller (see \secref{sec:controller-synthesis}).
		Appropriate modeling choices for the dynamics of the auxiliary variables $\ytrax$ and $\psitrax$ are based on the track curvature and the current velocity \cite{Morari.2010} such that
		\begin{align}
			\dot{\psi}_{\ur{TrAx}} =  {}& \dot{\psi}_{\ur{Ax}} - \dot{\psi}_{\ur{Tr}}, \label{eq.psi_trax}\\
			\dot{y}_{\ur{TrAx}} = {}& \dot{x} \sin \psitrax. \label{eq.y_trax}
		\end{align}
		Please note, by assuming \eqn{eq.y_trax} the lateral dynamics of the model are fully determined by $\dot{x}$ and $\psi_{\ur{Ax}}$.\\
		Regarding the dependent variables $z_{\ur{TrAx}}$ and $\varphi_{\ur{TrAx}}$, conic wheels on a line-shaped rail are assumed based on \cite{Jaschinski.1990,AndreasHeckmann.2016}, which leads to the approximations 
		\begin{align}\label{eq.z_and_phi}
			\varphi_{\ur{TrAx}} = {}& - \Gamma \ytrax + \varphi_{\ur{Tr}}, \\
			z_{\ur{TrAx}} = {}& \frac{\delta_{0} b}{2} \left(\frac{1}{\cos \psitrax} - 1 \right)- \Gamma \ytrax^{2} - r_0,
		\end{align}
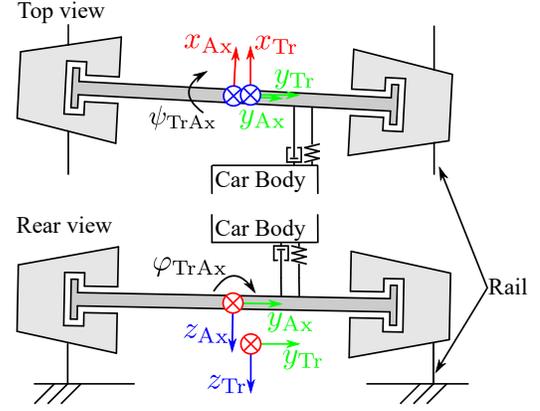
		\newfig[t]{width=6.8cm}{figure_3}{Simplified model of a railway running gear with \ac{irw} for use in \ac{mpc}.}\\
		\noindent in which $\Gamma=\frac{\tan \delta_0}{b/2 - r_0 \tan \delta_0}$ is a geometrical parameter, $b$ is the track gauge, $r_0$ is the nominal wheel radius and $\delta_0$ is the contact and cone angle, respectively, in the simplified model. The current wheel radii $r_j$ and lateral distances between wheel carrier center and rails $y_j$ can be inferred from the above relationships as well.\\
		Despite extensive research, the reliable online estimation of parameters $\cali{C}_{\ur{Tr}}$ describing the exact adhesion-slip characteristic remains difficult \cite{Shrestha.2019}. 
		Therefore, we assume that an accurately parameterized model of the adhesion-slip characteristic is not available, and the contact forces cannot be modeled realistically. 
		In this light, we simplify the model further by eliminating the wheel rotations from the generalized coordinate vector to reduce model complexity and computational effort in \ac{mpc}. 
		This is done by assuming ideal rolling, which results in a direct coupling of the wheel rotational velocities with the longitudinal running gear velocity $\dot{x}$ and the rotational velocities of running gear $\dot{\psi}_{\ur{Ax}}$ and track $\dot{\psi}_{\ur{Tr}}$, \ie
		\begin{subequations}\label{eq.wheel_speeds}
			\begin{align}
				\omega_{\ur{ri}} = {} & -\frac{1}{r_{\ur{ri}}} \left( \dot{x} - y_{\ur{ri}} \left( \dot{\psi}_{\ur{Tr}} + \dot{\psi}_{\ur{Ax}} \right) \right), \\
				\omega_{\ur{le}} = {} & -\frac{1}{r_{\ur{le}}} \left( \dot{x} + y_{\ur{le}} \left( \dot{\psi}_{\ur{Tr}} + \dot{\psi}_{\ur{Ax}} \right) \right).
			\end{align}
		\end{subequations}
		In order to apply the \lagrange formalism, expressions for the kinetic energy $E_{\ur{T}}$, the potential energy $E_{\ur{V}}$ and the dissipation function $E_{\ur{D}}$ of the running gear system can be obtained, which read
		\begin{subequations}\label{Lagrange}
			\begin{align}
				\begin{split}\label{Lagrangea}
					E_{\ur{T}} ={}& \frac{1}{2} \left(  m_{x} \dot{x}^2 + m \left( \dot{y}_{\ur{TrAx}}^2 + \dot{z}_{\ur{TrAx}}^2 \right) \right) \\ 
					+ & \frac{1}{2} \left( J_{\ur{Ax}_z} \dot{\psi}_{\ur{Ax}}^2 + J_{\ur{Ax}_x} \dot{\varphi}_{\ur{TrAx}}^2 \right)\\
					+ & \frac{1}{2} \left( \bi{\omega}_{\ur{ri}}\tr + \bi{\omega}_{\ur{le}}\tr \right)\bi{J}_{\ur{W}} \left( \bi{\omega}_{\ur{ri}} + \bi{\omega}_{\ur{le}} \right) ,
				\end{split}
			\end{align}
		\end{subequations}
		\begin{subequations}\label{Lagrange3}
			\begin{align}
				E_{\ur{V}} = {}& \frac{1}{2} \left( k_{\ur{s}_z} \left( \psi_{\ur{Ax}} - \psi_{\ur{CB}} \right)^2 + k_{\ur{s}_x} \varphi_{\ur{TrAx}}^2\right) - m g z_{\ur{TrAx}}, \label{Lagrangeb}\tag{$13\ur{b}$}\\
				E_{\ur{D}} = {}& \frac{1}{2} \left( k_{\ur{d}_z} \left( \dot{\psi}_{\ur{Ax}} - \dot{\psi}_{\ur{CB}} \right)^2 + k_{\ur{d}_x} \dot{\varphi}_{\ur{TrAx}}^2\right). \label{Lagrangec}\tag{$13\ur{c}$}
			\end{align}
		\end{subequations}\addtocounter{equation}{-1}
		
		\noindent The vectors $\bi{\omega}_j$ and the matrix $\bi{J}_{\ur{W}}$ describe the absolute rotation and the moment of inertia of the wheels to account for gyroscopic effects. The mass $m$ is the joint mass of the wheel carrier and the wheels. Similarly, $J_{\ur{Ax}_x}$ and  $J_{\ur{Ax}_z}$ are the residual moments of inertia regarding a roll and a yaw motion in $\CF{Ax}$. The mass $m_x = m + m_{\ur{CB}}/2$ accounts for the increased inertia in longitudinal direction, since half the car body mass needs to be accelerated by one running gear. The moment of inertia of the wheels about their primary rotation axis is denoted $J_{\ur{W}_y}$.\\
		The parameters $k_{\ur{s}_x}$, $k_{\ur{s}_z}$, $k_{\ur{d}_x}$ and $k_{\ur{d}_z}$ are spring and damping constants, respectively. Please note, the yaw spring and damping moments do not occur due to relative yaw angles and rates between $\CF{Tr}$ and $\CF{Ax}$ but with respect to the car body. The corresponding quantities of the car body $\psi_{\ur{CB}}$ and $\dot{\psi}_{\ur{CB}}$ can be approximated by simple geometrical considerations. Using available track data, the mean track yaw angle and rate between front and rear running gear can be computed by 
		\begin{align}\label{track_dep_variables}
			\psi_{\ur{CB}} = {}& \frac{\psi_{\ur{Tr}}(p) + \psi_{\ur{Tr}}(p - L_{\ur{CB}})}{2}, \\
			\dot{\psi}_{\ur{CB}} = {}& \frac{\dot{\psi}_{\ur{Tr}}(p) + \dot{\psi}_{\ur{Tr}}(p - L_{\ur{CB}})}{2},
		\end{align}
		where $L_{\ur{CB}}$ denotes the distance between front and rear running gear.\\
		Applying the simplifications \eqn{eq.y_trax} through \eqn{eq.wheel_speeds}, the expressions \eqn{Lagrange} become solely dependent on a reduced vector of generalized coordinates $\bi{q}_{\ur{red}}\tr~=~\left[x,\psi_{\ur{Ax}}\right]$. The corresponding \lagrange formalism reads
		\begin{subequations}\label{Lagrange2}
			\begin{align}
				&\cali{L} = {} E_{\ur{T}} - E_{\ur{V}} \label{Lagranged}, \\
				& \frac{\ur{d}}{\ur{d}t}  \left( \frac{\partial \cali{L}}{\partial \dot{\bi{q}}_{\ur{red}}} \right) - \frac{\partial \cali{L}}{\partial \bi{q}_{\ur{red}}} + \frac{\partial E_{\ur{D}}}{\partial \dot{\bi{q}}_{\ur{red}}} = \bi{F}_{\ur{gen}}, \label{Lagrangee}
			\end{align}
		\end{subequations}
		where the constraints following from the above assumptions are considered directly in the generalized coordinates $\bi{q}_{\ur{red}}$ and generalized forces $\bi{F}_{\ur{gen}}$ and not utilizing \lagrange multipliers.\\
		Due to the assumption of ideal rolling, the motor torques $\tau_{j}$ act immediately on the reduced generalized coordinates $\bi{q}_{\ur{red}}$. Therefore, the generalized forces $\bi{F}_{\ur{gen}}$ can be approximated for small angles $\psitrax$ by
		\begin{equation}\label{gen_forces}
			\bi{F}_{\ur{gen}} = - \begin{bmatrix}
				\frac{\tau_{\ur{le}}}{r_{\ur{le}}} + \frac{\tau_{\ur{ri}}}{r_{\ur{ri}}} \\ \frac{y_{\ur{le}} \tau_{\ur{le}}}{r_{\ur{le}}} - \frac{y_{\ur{ri}} \tau_{\ur{ri}}}{r_{\ur{ri}}}		
			\end{bmatrix}.
		\end{equation}
		To formulate a state-space model, the longitudinal and yaw system dynamics in $x$ and $\psi_{\ur{Ax}}$ are obtained from \eqn{Lagrangee}, respectively. In addition, \eqn{eq.psi_trax} and \eqn{eq.y_trax} are appended to the state vector. This allows for path following such that the lateral displacement $\ytrax$ and the relative yaw angle $\psitrax$ between running gear and track center line can be controlled actively. Besides, track preview information can be incorporated. Thus, a control-oriented continuous-time state-space model with 
		\begin{align}
			\bi{x}\tr =& \begin{bmatrix}
				x & \psi_{\ur{Ax}} & \dot{x} & \dot{\psi}_{\ur{Ax}} & \ytrax & \psitrax
			\end{bmatrix},\\
			\bi{u}\tr =& \begin{bmatrix}
				\tau_{\ur{ri}} & \tau_{\ur{le}}
			\end{bmatrix}, \\
			\theta =& \begin{Bmatrix}
				\cali{D}_{\ur{Tr}} & m_{\ur{CB}}
			\end{Bmatrix}, \label{eqn.parameters}
		\end{align}
		can be defined. Using a discretization method, \eg the \euler algorithm, the desired discrete-time representation $\bi{f}_{\ur{d}}$ of the nonlinear state-space model is obtained.\\
		Please note, the derived control-oriented prediction model is meant for \ac{mpc}-based mechatronic guidance. To obtain a computationally efficient formulation, simplifying assumptions, such as conic wheels and ideal rolling, are used. As the simulation results in \secref{sec:experimental-results} show, these assumptions are valid in the considered scenarios. In other control contexts, conditions may differ, making model adjustments necessary. 
		\newfig[b]{width=0.48\textwidth}{figure_4}{Schematic of the \ac{mpc}-based integrated control system for railway running gears.}
		\section{Controller Synthesis}\label{sec:controller-synthesis}
		The overall controller is presented in \fig{figure_4} and consists of three subsystems, the lateral controller, the longitudinal controller, and the integration rule.
		As can be seen, the desired lateral position $y^{*}$ and the demanded traction or braking force $F_{x}^{*}$ of the running gear are set points for the control system.
		It is assumed that these values are provided by a high-level system controlling all running gears holistically in accordance with the project specification of \ac{ngt}.\\
		In detail, the lateral controller employs the state-space model from \secref{sec:control-oriented-model} in an \ac{mpc} scheme to determine a differential control torque $\Delta u$. Its objective is to steer the running gear such that the actual lateral displacement $\ytrax$ approaches the set point $y^{*}$. Inputs for the longitudinal controller are the demanded traction or braking force $F_{x}^{*}$ as well as information about the momentary adhesion and slip values between wheels and rails. It identifies the segment of the adhesion-slip characteristic, in which the system operates, and computes a base control torque $u^\ur{d}$ to satisfy $F_{x}^{*}$ while preventing locking or skidding of the wheels. The controller operates without relying on a specific contact model of the wheel-rail contacts. The two control torques are combined by an integration rule and given to the motors.\\
		A main advantage of the proposed \ac{mpc} controller is that it can consider known track geometry (\eg from a database) and known set point trajectories inherently in the prediction model. Here, we exploit this idea to show its effectiveness for refining the lateral guidance of railway running gears with \ac{irw}. The core methods are developed in this contribution, assuming the availability of a database with geometric data about the track and accurate train localization along the track.\\
		The first assumption is valid since geometric track data is usually available from track network operators \cite{DBNetzAG.2023}. The second assumption is not restrictive as well, since promising results have been obtained with novel localization techniques. For instance, a novel method for train localization presented in \cite{Siebler.2020,Siebler.2023} makes use of local variations of the Earth's magnetic field to supplement conventional \acp{gnss} and increase the reliability of train localization. The method employs a particle filter that performs sensor fusion, taking into account magnetometer measurements and a model of the train dynamics. Loosely speaking, the accurate position along the track $p$ is inferred by comparing the currently measured magnetic field to a previously measured magnetic footprint of the track \cite{Siebler.2020,Siebler.2023}. A main advantage is the availability in areas without network connection (\eg tunnels), increasing reliability. First experimental results indicated an accuracy of about $2\,\ur{m}$ \cite{Siebler.2023}.\\ 
		Further, the integrated control system is provided with measurements of states $\hat{\bi{x}}$, adhesion $\hat{f}_{x_j}$, and slip $\hat{s}_{x_j}$. 
		A main objective of the current contribution is to investigate \ac{mpc} approaches in railway running gears. In this context, estimator design is out of scope, and the \ac{mbs} outputs are used as direct measurements for feedback control. 
		However, powerful nonlinear \acp{kf} do exist for comparable applications \cite{Keck.2021,Schwarz.2021}, and an estimator-controller combination has to be employed in future implementations. This is of importance since the estimator can be used to obtain state information even if certain variables are difficult or costly to measure, as is the case with states $\ytrax$ and $\psitrax$ in the current setting \cite{Keck.2021,Schwarz.2021}. 
		In addition, the fusion of model and sensor information can reduce noise and improve physical consistency.\\ 
		After some general remarks on the control framework and related assumptions, the three subsystems are explained in more detail subsequently. First, the adhesion controller is explained in \secref{sec:adhesion-control}. Second, the \ac{mpc}-based lateral guidance controller is proposed. Last, an integration system that combines $u^{\ur{d}}$ and $\Delta u$ is introduced based on \cite{Schwarz.2022b}.
		
		\subsection{Adhesion Control}\label{sec:adhesion-control}
		The adhesion controller should manage regular braking and acceleration scenarios as well as situations in which a maximum adhesion between wheel and rail is demanded. The latter occurs mostly during emergency braking or when adhesion conditions between wheels and rails are very poor. For these cases, a maximum-seeking adhesion controller, which does not rely on a specific parametrized contact model, has been proposed in \cite{Schwarz.2021}. In this contribution, the controller is revisited briefly and developed further for general autonomous operation without knowledge of the adhesion-slip characteristic. The controller is based on a sliding mode approach and approximates the switching function $\sigma = \partial f_{x}/\partial s_{x}$ using
		\begin{equation}\label{eqn.MAC_switching_function}
			\bar{\sigma} = \dot{\hat{f}}_{x} \dot{\hat{s}}_{x},
		\end{equation}
		with estimates of longitudinal adhesion $f_{x}$, and slip $s_x$. The estimates can be obtained from nonlinear estimation algorithms designed in \cite{Schwarz.2019c}, which are assumed to be given here. For the current application, the control law is chosen such that the sliding mode approach determines a motor torque change $\delta u$, which is added to the absolute torque in every time step, \ie
		\begin{align}\label{eqn.MAC_control_law}
			u^{\ur{d}}_{k} &= u^{\ur{d}}_{k-1} + \delta u, \\	
			\delta u &= p_1 \ur{sgn} \left(\bar{\sigma}\right),
		\end{align}
		where $p_1$ is a control parameter to be chosen sufficiently high.
		The resulting control behavior is illustrated in \fig{figure_5} on the right side, where the set point adhesion $f_{x_j}^{*}$ cannot be reached due to poor adhesion conditions (\ie $f_{x_{j}}^{*}>f_{x_{\ur{max}}}$). Nonetheless, the maximum possible adhesion $f_{x_{\ur{max}}}$ should be exploited. The control law causes the motor torque to
		\begin{itemize}[noitemsep,labelindent=\parindent,leftmargin=20pt,topsep=0pt]
			\item increase if the operating point is in segment (I) below its maximum adhesion value (\ie $\bar{\sigma}>0$), and to
			\item decrease as soon as the unstable segment (III) is entered (\ie $\bar{\sigma}<0$).
		\end{itemize}
		Therefore, the controller keeps oscillating around $f_{x_{\ur{max}}}$ and eventually converges to a point close to it. The detailed proof of convergence is conducted in \cite{Schwarz.2021}. It is based on \lyapunov arguments and a concavity assumption regarding the adhesion-slip characteristic in the considered region (\ie $\partial^2 f_{x}/\partial s_{x}^{2}<0$).\\
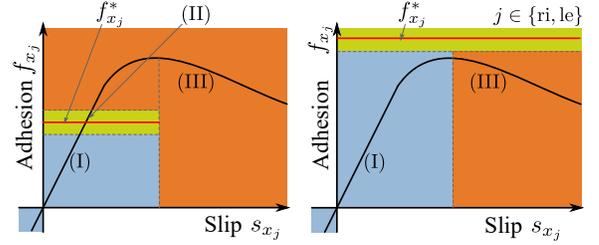
		\newfig{width=7.6cm}{figure_5}{Operational segments of adhesion control scheme for two scenarios: desired adhesion $f_{x_j}^{*}<f_{x_{\ur{max}}}$ (left) or $f_{x_j}^{*}>f_{x_{\ur{max}}}$ (right). Operation in segment (I): torque to be increased. Operation in segment (II): no torque change. Operation in segment (III): torque to be decreased.}
		Beyond the current state of research, we extend this maximum-seeking controller to general application in regular as well as poor adhesion situations, during both traction and braking. To this end, a control law for regular operation is devised following the lines of \cite{Schwarz.2021} and combining it with the above maximum adhesion control law \eqn{eqn.MAC_control_law}.\\
		Applying $\lyapunov$ theory to the regular braking case where the set point adhesion $f_{x}^{*}$ lies below the maximum value of the adhesion-slip characteristic (see \fig{figure_5}, left), we show that the control error
		\begin{equation}
			\nu = f_{x}^{*} - f_{x} \approx f_{x}^{*} - \hat{f}_{x} = \bar{\nu},
		\end{equation}  
		converges to zero under the control update law
		\begin{equation}\label{eq.RAC_law}
			\delta u = p_2 \ur{sgn} \left(\bar{\nu}\right),
		\end{equation}  
		where $p_2$ is a control parameter. To prove convergence, the estimated values are assumed to be exact, and hence the ideal error function $\nu$ is employed. A suitable candidate \lyapunov function is
		\begin{equation}\label{eqn.lyap}
			V(\bi{x}) = \frac{1}{2} \nu^2.
		\end{equation}  
		For this regular braking and acceleration case, the operating point on the adhesion slip characteristic locates in the micro slip regime. Thus, we model the longitudinal adhesion by linear Kalker theory \cite{KALKER.1982}, \ie
		\begin{equation}\label{eqn.kalkerx}
			f_x = ab G c_{11} s_x,
		\end{equation} 
		with the Kalker coefficient $c_{11}$, the shear modulus $G$, and the half axes of the contact ellipse $a$ and $b$.\\
		Further, if we consider the situation when driving at constant speed, constant adhesion forces between wheels and rails are needed to counteract resistance forces such as friction. This implies $\omega r_j > \dot{x}$, which gives a constant negative slip value using the definition
		\begin{equation}
			s_{x} = \frac{\dot{x} - \omega r_j}{\dot{x}}.\\
		\end{equation}
		Starting from the above stationary driving condition, an increase in the motor torque $\delta u$ leads to an increased rotational velocity of the wheel (\ie $\delta u~=~J_{\ur{W}_y} \dot{\omega}$), which causes a further decrease of the slip. This can be seen in the slip derivative
		\begin{subequations}
			\begin{align}
				\dot{s}_x &= \frac{\dot{x}\left(\ddot{x} - \dot{\omega} r_j\right) - \ddot{x}\left(\dot{x} - \omega r_j\right)}{\dot{x}^2}\\
				&= f(\bi{x}) + g(\bi{x}) \delta u,
			\end{align}
		\end{subequations}
		as well, where $f(\bi{x})=\frac{   \omega r_j   }{\dot{x}^2} \ddot{x}$ is bounded and $g(\bi{x})=-\frac{ r_j  }{\dot{x} J_{\ur{W}_y}}$ is negative. Since the velocity $\dot{x}$ is bounded, $g(\bi{x})\neq0$ holds.
		Hence, assuming a constant desired adhesion $f_x^{*}$, the time derivative of \eqn{eqn.lyap} reads
		\begin{subequations}
			\begin{align}
				\dot{V}(\bi{x}) &= \nu \dot{\nu}\\
				&= - \nu \frac{\partial f_{x} }{\partial s_{x}} \dot{s}_{x} \\
				&= \nu \left( ab G c_{11} \right) \left(f(\bi{x}) + g(\bi{x}) p_2 \ur{sgn} \left({\nu}\right) \right)  \\
				&= \nu c f(\bi{x}) +\nu c g(\bi{x}) p_2 \ur{sgn} (\nu)\\
				&= \nu c f(\bi{x}) +\left| \nu \right| c g(\bi{x}) p_2, \label{eq.lyap_proof}
			\end{align} 
		\end{subequations} 
		where $c$ is a positive constant. From \eqn{eq.lyap_proof}, it is easy to see that a sufficiently high $p_2$ can ensure negative definiteness of $\dot{V}(\bi{x})$, which renders the operation point $\nu = 0$ attractive. Under the standing assumptions, the control law \eqn{eq.RAC_law} causes the adhesion to converge to the set point $f_{x}^{*}$.\\
		Taking the above findings further, our adhesion control method employs additional switching functions to account for traction and braking as well as for the case when the system operates in the third quadrant of the adhesion-slip characteristic. Besides, a tolerance corridor around the demanded adhesion value $f_x^{*}$ avoids oscillations. Hence, the adhesion-slip diagram is divided into multiple subsegments, on which suitably chosen control design parameters $p_i$ determine $\delta u$. Computationally efficient logical statements (such as $\dot{\hat{f}}_x \dot{\hat{s}}_x>0$ or $f^{*}_{x}-\hat{f}_{x}>0$) help to determine the subsegment in which the system operates.
		
		\subsection{Nonlinear Model Predictive Control}\label{sec:model-predictive-control}
		\ac{mpc} is a control method in which an optimization problem is solved to determine the ``optimal'' control input sequence with respect to a cost function. The first instance of the sequence is applied to the system, and the optimization process is started again. The cost function is typically associated with the difference between the desired and the predicted future states over a prediction horizon $H$, when applying a certain control input sequence. For actual implementation, the prediction horizon is divided in $L$ discrete time steps with duration $T=H/L$.\\
		For mechatronic guidance of railway running gears, the use of \ac{mpc} is very appealing since known quantities (\eg set point trajectories or geometric track data) can be inherently considered in the prediction horizon for precise and reliable control. In this light, the current contribution is the first to exploit \ac{mpc} and its conceptual advantages for lateral control of railway running gears with \ac{irw}. The proposed \ac{mpc} scheme takes the desired lateral position of the running gear in the track as a set point, which corresponds to $\ytrax$ in the control-oriented model (see \fig{figure_4}). Besides, the track geometry is considered as varying parameters in accordance with \eqn{eqn.parameters}. Additionally, the estimated states and the currently desired mean torque $u^{\ur{d}}$ are made available to the \ac{mpc}. The latter is considered to improve prediction precision and hence control performance. It is worth mentioning that not only momentary set points and track data can be used. In fact, whole sequences, \eg dependent on the track coordinate, can be considered in the lateral controller.\\
		The applied \ac{mpc}-scheme is based on the control-oriented model introduced in \secref{sec:control-oriented-model}. The model is used to predict future states $\ps[]{x}{0,L}$ if a certain input sequence is applied. To retain a concise, yet brief description, a sequence $\left\{ \bi{z}_k \right\}_{k=v}^{w}$ is denoted $\bi{z}_{\left[v,w\right]}$. Note further, predicted quantities are shown with a bar and a quantity followed by $(t)$ is processed in the algorithm at time $t$.\\
		Following standard \ac{mpc} notation and using \eqn{eq.discr_state_space1}, the \ac{nmpc} optimization problem reads
		\begin{subequations}\label{eq.constraints}
			\begin{align}
				\min_{\Delta \bar{u}_{[0,L-1]}(t)} &\sum_{k=0}^{L-1} T \norm{\bar{\bi{x}}_k(t) - \bi{x}_k^\ur{d}(t)}_{\bi{Q}}^2  +T\norm{\Delta \bar{{u}}_k(t)}_{{R}}^2 \nonumber\\
				& \qquad \qquad \qquad \qquad \qquad \qquad + J_\ur{term}, \\
				\ur{s.t.} \: & \bar{\bi{x}}_{k+1}(t) = \bi{f}_{\ur{d}} \left(\bar{\bi{x}}_k(t),\bar{\bi{u}}_{k}(t), \theta \right),  \label{eq.constraintsa}\\
				& \bar{\bi{x}}_{0}(t) = \hat{\bi{x}}_{t}, \\
				& \bar{\bi{u}}_{k}(t) = \begin{bmatrix}
					u_k^{\ur{d}}(t) + \Delta \bar{u}_k(t)\\ u_k^{\ur{d}}(t) - \Delta \bar{u}_k(t)
				\end{bmatrix}, \\
				& \bar{\bi{u}}_{[0,L-1]}(t) \in \bi{\cali{U}}, \\
				& \bar{\bi{x}}_{[0,L]}(t) \in \bi{\cali{X}}, \qquad \forall k \in [0,L-1],
			\end{align}
		\end{subequations}
		where $\norm{\bi{v}}_{\bi{M}}^2 = \bi{v}\tr \bi{M} \bi{v}$. $\bi{\cali{X}}$ and $\bi{\cali{U}}$ are the static state and input constraints, respectively. $\bi{Q}$ and $R$ are cost weighting parameters of suitable dimensions. $\hat{\bi{x}}_t$ is the estimated state vector at time $t$ and $J_{\ur{term}}$ is an appropriate additive term to approximate the terminal cost. In the current implementation, a suitable choice which facilitates fast convergence to the desired set point $y^{*}$ without static terminal constraints is
		\begin{equation}\label{eq.terminal_cost}
			J_{\ur{term}} = T \norm{\bar{\bi{x}}_L(t) - \bi{x}_L^\ur{d}(t)}_{\bi{Q}_L}^2,
		\end{equation}
		where $\bi{Q}_L = \bi{Q} \cdot \ur{diag}\left(0,0,0,0,q,q\right)$ and $q$ is a control design parameter.\\
		The desired state sequence considered in the \ac{mpc} horizon
		\begin{equation}
			\bi{x}_{k}^{\ur{d}}(t)\tr = \begin{bmatrix}
				x_{k}^{\ur{d}} & \psi_{\ur{Ax}_{k}}^{\ur{d}} & 
				\dot{x}_{k}^{\ur{d}} & \dot{\psi}_{\ur{Ax}_{k}}^{\ur{d}} & 
				y_{\ur{TrAx}_k}^{\ur{d}} & \psi_{\ur{TrAx}_k}^{\ur{d}} 
			\end{bmatrix},
		\end{equation}
		is dependent on the momentary position and velocity along the track. For an efficient calculation, an approximation of the future track position and velocity is made over the prediction horizon. It is built from the desired lateral displacement sequence $y^{*}$, the track preview parameters $\theta$, and the currently estimated state vector $\hat{\bi{x}}_t$. For instance, the estimated state vector $\hat{\bi{x}}_t$ provides the initial position $\hat{x}_t$ of the running gear, and the approximate future positions are computed using current velocity $\hat{\dot{x}}_t$ and desired base torque $u_k^{\ur{d}}$. For track preview, the track geometry parameters $\theta$ are found by interpolation at the approximated future positions. In detail,
		\begin{subequations}\label{eq.des_states}
			\begin{align}
				&x_{k}^{\ur{d}} &&= \hat{x}_t + \hat{\dot{x}}_t \: T k + \frac{u_k^{\ur{d}}}{r_0 m_x} \frac{\left(T k \right)^2}{2},\\
				&\psi_{\ur{Ax}_{k}}^{\ur{d}} &&= \psi_{\ur{TrAx}_k}^{\ur{d}} + \psi_{\ur{Tr}}(x_{k}^{\ur{d}}),\\
				&\dot{x}_k^{\ur{d}} &&= \hat{\dot{x}}_t + \frac{u^{\ur{d}}_k}{r_0 m_x} T k,\\
				&\dot{\psi}_{\ur{Ax}_{k}}^{\ur{d}} &&= \dot{\psi}_{\ur{TrAx}_k}^{\ur{d}} + \dot{\psi}_{\ur{Tr}}(x_{k}^{\ur{d}}),\\
				&y_{\ur{TrAx}_k}^{\ur{d}} &&= y^{*}(x_{k}^{\ur{d}}),\\
				&\psi_{\ur{TrAx}_k}^{\ur{d}} &&= \frac{\dot{y}_{\ur{TrAx}_k}^{\ur{d}}}{\dot{x}_k^{\ur{d}}}.
			\end{align}
		\end{subequations}
		The same approximation is used for determination of the track preview in the system model.
		\subsection{Linear Time-Variant Model Predictive Control}\label{sec:ltv-model-predictive-control}
		In the previous section, a novel \ac{nmpc}-based running gear controller is developed. It exploits the conceptual advantage of \ac{mpc} to consider known quantities (\eg set point trajectories or geometric track data) in a prediction horizon. A known drawback of \ac{mpc} is a high computational burden. Thus, besides the main objective to show the general potential of \ac{mpc} for railway running gear control, we present a simplified version of the controller, which is computationally less demanding.\\
		To this end, the control-oriented system model \eqn{eq.discr_state_space1} is linearized in a centered riding position around the approximated future track position and velocity according to
		\begin{align}\label{eq.des_states2}
			\bi{A}_k &= \left. \frac{\partial \bi{f}_{\ur{d}}}{\partial \bi{x}} \right\rvert_{\bi{x}=\bi{x}_{\ur{lin}_k}, \bi{u}=\bi{u}_{\ur{lin}_k}, \theta=\theta_{\ur{lin}_k}},\\	
			\bi{B}_k &= \left. \frac{\partial \bi{f}_{\ur{d}}}{\partial \bi{u}} \right\rvert_{\bi{x}=\bi{x}_{\ur{lin}_k}, \bi{u}=\bi{u}_{\ur{lin}_k}, \theta=\theta_{\ur{lin}_k}},
		\end{align}
		\begin{align}\label{eq.des_states3}
			\text{where } \bi{x}_{\ur{lin}_k} &= \begin{bmatrix}
				x^{\ur{d}}_k & 0 & \dot{x}^{\ur{d}}_k & 0 & 0 & 0
			\end{bmatrix}\tr,\\
			\bi{u}_{\ur{lin}_k} &= \begin{bmatrix}
				u^{\ur{d}}_k & u^{\ur{d}}_k
			\end{bmatrix}\tr,\\
			\theta_{\ur{lin}_k} &= \theta(x^{\ur{d}}_k, \dot{x}^{\ur{d}}_k),
		\end{align}
		and $\theta(x^{\ur{d}}_k, \dot{x}^{\ur{d}}_k)$ denotes the set of variable parameters, dependent on the position and velocity forecasts.
		The dynamic constraint \eqn{eq.constraintsa} is reformulated using the \ac{ltv} system model
		\begin{equation}\label{eq.lin_model}
			\bar{\bi{x}}_{k+1}(t) = \bi{A}_k \bar{\bi{x}}_k(t) + \bi{B}_k \bar{\bi{u}}_{k}(t),
		\end{equation}
		to obtain an \ac{ltvmpc} problem. The remaining parts of the controller are unchanged.
		
		\subsection{Integration Approach}\label{sec:integration-approach}
		The employed integration approach flexibly allocates a control torque to each of the subsystem controllers in lateral and longitudinal direction. In contrast to existing implementations \cite{Schwarz.2022b}, the use of \ac{mpc} allows for an explainable and smooth shift of the control objective between lateral and longitudinal emphasis. This is done by tuning the parameters $\bi{Q}$ and $R$ in the \ac{mpc} scheme. Loosely speaking, the cost for allocating a certain lateral control torque $\Delta u_k(t)$ and for deviating from the set points $\bi{x}_k^\ur{d}(t)$ can be adjusted by the operator in an explainable fashion.\\
		Following the approach in \cite{Schwarz.2022b}, no fixed torque values are allocated to the longitudinal and lateral controller. Rather, the output torques are determined by the integration rule
		\begin{align}\label{eq.integration_rule}
			{\tau_{\ur{long}_k}} =& \begin{cases}
				{\min \left(u^{\ur{d}}_k,\tau_{\ur{max}}-\left|\Delta u_k\right| \right),}&{\text{if}}\ u^{\ur{d}}_k>0,\\ 
				{\max \left(u^{\ur{d}}_k,\tau_{\ur{min}}+\left|\Delta u_k\right| \right),}&{\text{otherwise,}} 
			\end{cases}\\
			{\bi{u}_k} =& \begin{bmatrix}
				\tau_{\ur{long}_k} + \Delta u_k \\
				\tau_{\ur{long}_k} - \Delta u_k
			\end{bmatrix}
		\end{align}
		where $\tau_{\ur{max}}$ and $\tau_{\ur{min}}$ are the maximum and minimum possible wheel motor torques, respectively.\\
		For the adhesion controller, it is crucial to react quickly if adhesion conditions worsen abruptly. Therefore, the adhesion control operates at a high frequency, \eg $1000\, \ur{Hz}$. The lateral controller, however, solves an optimization problem repetitively. In this light, computational resources do limit the maximum frequency at which the \ac{mpc}-scheme can calculate control outputs. The integration rule considers this frequency mismatch and ensures the commanded differential torque $\Delta u$, which is determined by the \ac{mpc}, even if $u^{\ur{d}}$ changes sharply.
		
		\section{Multi-Body Simulation Results}\label{sec:experimental-results}
		To show the effectiveness of the proposed integrated controller, extensive simulations are conducted. The simulation environment is described in \secref{sec:simulation-environment} and the corresponding evaluation tracks are presented in \secref{sec:test-scenarios}. 
		Since the proposed integrated controller needs to satisfy both, lateral and longitudinal control objectives, different evaluation scenarios are considered. First,
		the capability of the mechatronic guidance subsystem is tested in challenging scenarios with emphasis on lateral railway running gear control (\secref{sec:evaluation_lateral}). Second, an evaluation with primarily longitudinal control objectives is conducted. Third, the integrated controller is tested in a scenario that demands both, high lateral and longitudinal control action. In this evaluation, we show that the proposed controller allows for a custom and smooth shift between different control objectives (\secref{sec:evaluation_longitudinal}). The detailed evaluation procedure and performance criteria are explained in the corresponding evaluation sections.\\
		To show the effectiveness of the proposed integrated control scheme, the results are compared to a state-of-the-art controller. It is worth mentioning that few control systems for mechatronic guidance of railway running gears with driven \ac{irw} exist, and even fewer controllers allow for integrated longitudinal and lateral running gear control. To the best of our knowledge, there are no other \ac{mpc}-approaches for railway running gears with \ac{irw}. 
		Beyond the current state of research, we exploit the predictive nature of \ac{mpc} to incorporate preview information about the track into control. In this light, \ac{mpc} has a conceptual advantage over existing running gear controllers, and few methods for direct comparison exist.\\ 
		Therefore, we employ the advanced \ac{ndi} \cite{Ducard.2008} controller presented in \cite{Grether.2018,Heckmann.2020} for comparison. In the literature, this technique is an established control method for railway running gears with \ac{irw}, making it a valid choice for performance comparison. In \cite{Grether.2018,Heckmann.2020}, it follows a feedback linearization approach and features a simplified nonlinear inverse system model of the running gear to accomplish mechatronic guidance. For further details, the reader is referred to \cite{Grether.2018,Heckmann.2020}.\\
		Since the corresponding plant models are known, the parameters of the \ac{mpc}-model are assumed to be known as well and no parameter identification is performed. The lateral control design parameters $\bi{Q}$, $R$, $q$, the prediction horizon length $H$, and time step $T$ are tuned using a global optimization algorithm. In detail, a \ac{pso} minimizes a cost function along the design parameters \cite{Kennedy.1995}. The cost function is associated with the difference $\ytrax - y^{*}$ and the magnitude of the differential torque $\Delta u$. Multiple velocities and track geometries are considered. The longitudinal control design parameters $p_i$ are tuned heuristically.
		
		\subsection{Simulation Environment}\label{sec:simulation-environment}
		The simulations are performed with a detailed model of a single high-speed train car and using the \ac{mbs} software \simpack. The controller implementation is based on \matlab and \simulink. In the experiments, \simulink and \simpack communicate via TCP/IP protocol at $1 \,\ur{kHz}$ in a co-simulation environment.\\
		The \ac{nmpc}-scheme is implemented with the nonlinear optimization toolbox \casadi \cite{Andersson.InPress2018}, which provides a comprehensive framework for convenient set-up of \ac{nmpc} problems. For solving of nonlinear optimization problems, the package \textsc{Ipopt} is used, which implements an interior-point method for nonlinear programming \cite{Wachter.2006}.\\
		For qualitative comparison of computational effort, computation times are given. The calculations are performed with an \intelcore i7-6700K CPU @ 4.00GHz, 16 GB RAM.
		
		\subsection{Test Tracks}\label{sec:test-scenarios}
		Since the application of \ac{mpc} in running gear control is new, our work focuses mainly on methodological aspects rather than a ready-to-use implementation. Therefore, control-oriented evaluation tracks and set point sequences $y^*$ are used to scrutinize and compare different core concepts. In this context, the evaluation tracks are geometrically ideal without any irregularities.\\
		On the one hand, challenging curve entering scenarios at different velocities are considered. In each scenario, the vehicle starts on a straight track, passes a clothoid and enters a curve with a constant radius. To show the effectiveness of the proposed schemes, the evaluation tracks $\cali{T}_{1}$ to $\cali{T}_{4}$ are designed for different velocities from $40$ to $400 \,\ur{km/h}$ and featuring different unbalanced lateral accelerations between $0$ and the operationally allowed maximum of $0.65 \,\ur{m/s^2}$. The curve radii are chosen comparable to previous publications \cite{Kurzeck.2014}, and the superelevation is hence a dependent quantity\footnote{The superelevation $l_{\ur{sup}}$ is deduced from the fixed values for curve radius $R$, velocity $v$ and design unbalanced lateral acceleration $a$ by $l_{\ur{sup}}~=~b\tan\varphi_{\ur{Tr}}$ and $\varphi_{\ur{Tr}} = \arcsin \frac{v^2/R - a}{g}$.\label{fnlabel}}. An overview of the scenarios can be found in \tab{Tracks}. In addition, the curve radius of track $\cali{T}_1$ is chosen such that the geometrically required yaw angle between wheel carrier and car body in perfect curving is at $80 \%$ of the mechanical yaw end stops. In this context, $\cali{T}_1$ is designed to test control performance in a very narrow curve, requiring high control effort.\\
		In addition, a further straight track $\cali{T}_{5}$ is considered for evaluation of methodological features and control performance. On this track, controllers can be tested independently of effects due to curving.
		\newtab[tb]{rlrlrl}{
			\hhline{======}
			& & & & & \\
			& & & & & \\
			& & & & & \\
			\myrotcell{Index $i$} & 
			\myrotcell{\makecell{Track \\shape}} & 
			\myrotcell{\makecell{Design \\velocity}} & \myrotcell{\makecell{Design lat. \\acceleration}} & \myrotcell{\makecell{Curve \\radius}}  & \myrotcell{\makecell{Super-\\elevation\footref{fnlabel}}}  \\
			& & $[\ur{km/h}]$ & $[\ur{m}/\ur{s}^2]$ & $[\ur{m}]$ & $[\ur{m}]$ \\
			\hline
			$1$ & straight-clothoid-curve & $40$ & $0$ & $175$ & $0.108$ \\
			$2$ & straight-clothoid-curve & $160$ & $0.2167$ & $1500$ & $0.168$ \\
			$3$ & straight-clothoid-curve & $280$ & $0.4333$ & $4250$ & $0.151$ \\
			$4$ & straight-clothoid-curve & $400$ & $0.6500$ & $8500$ & $0.123$ \\
			$5$ & straight  & $0$--$400$ & $0$ & $\infty$ & $0$ \\
			\hhline{======}}
		{Tracks}
		{Evaluation tracks $\cali{T}_{i}$ for \simpack{} multi-body co-simulation}
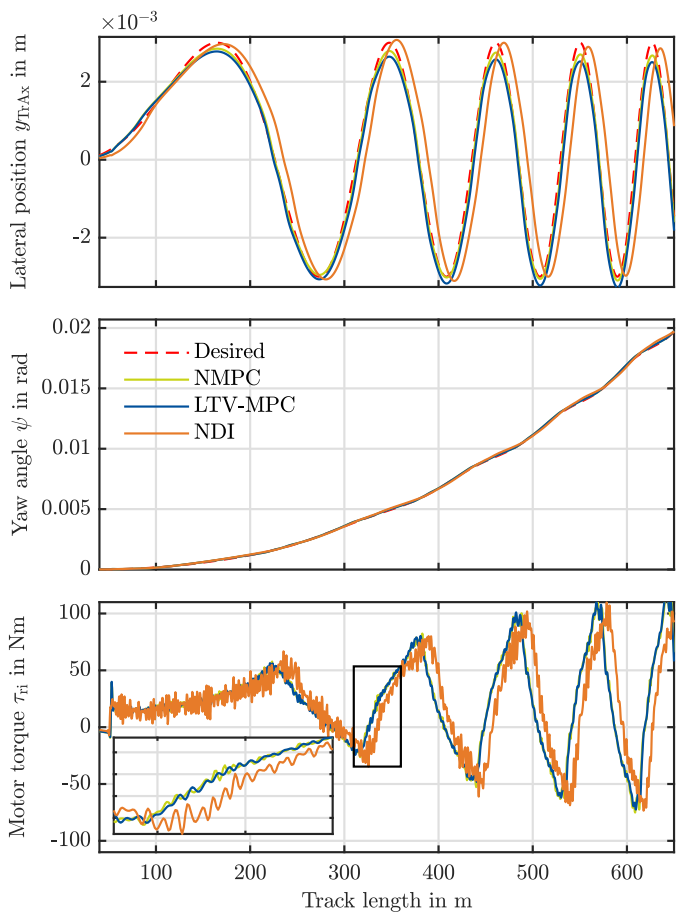
		\newfig[t]{width=0.49\textwidth}{figure_6}{Co-simulation result, evaluation track $\cali{T}_{3}$, $280 \,\ur{km/h}$, good adhesion conditions, $F_{x}^{*}=0$. All controllers allow for following of track path and set point trajectory. \ac{mpc} schemes can take future values into account, which reduces delay and hence tracking error. The results suggest that \ac{mpc} is well-suited to counteract highly dynamic, realistic lateral track irregularities in future implementations.}
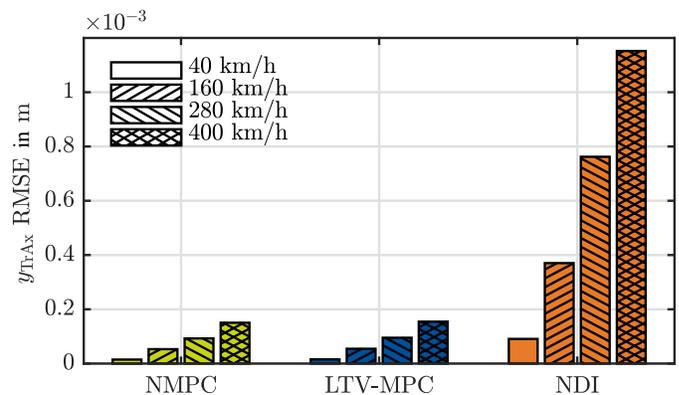
		\newfig[t]{width=0.49\textwidth}{figure_7}{Co-simulation result, $1000 \,\ur{m}$ of evaluation track $\cali{T}_5$, good adhesion conditions, $F_{x}^{*}=0$, sine with period $150\ur{m}$ as $y^{*}$ set point. The results indicate the superiority of \ac{mpc}-based lateral control in terms of lateral position tracking for a wide velocity range.}
		\subsection{Evaluation of Lateral Control}\label{sec:evaluation_lateral}
		In a first evaluation scenario (see \fig{figure_6} and \fig{figure_7}), the capability of active lateral guidance (\ie a dynamic desired lateral displacement $y^{*}$ has to be tracked) in challenging curve entering situations at velocities up to $400 \, \ur{km/h}$ is evaluated. The control performance criterion is the \ac{rmse} between the desired lateral displacement $y^{*}$ and the actual lateral displacement $\ytrax$ of the running gear with respect to the track center line. The results are compared to the state-of-the-art \ac{ndi} approach (see beginning of \secref{sec:experimental-results}).\\
		A specific example for qualitative analysis of this scenario is shown in \fig{figure_6}. It depicts the control behavior when riding on the curved track $\cali{T}_3$ without traction or braking (\ie $F_x^{*}=0$). 
		A variable set point trajectory of the lateral position $y^{*}$ in the form of a sine sweep is imposed. This set point trajectory is relevant since highly dynamic lateral track irregularities do occur on real tracks. Thus, accurate lateral control at high frequencies is crucial to avoid flange contact between wheels and rails. In addition, following a desired lateral position in the track allows for active planning of wheel wear and the resulting worn wheel geometry.\\
		Looking at the beginning of the clothoid at $25 \,\ur{m}$, it is visible that all controllers cope well with the change in superelevation, as they immediately counteract the gyroscopic effects by an adequate motor torque. The middle plot shows the absolute yaw angle and hence illustrates the overall path-tracking ability. In addition to following the clothoid geometry, it is possible to ensure the desired lateral position within the track.\\
		In \fig{figure_6}, the \ac{mpc}-based controllers show a better accuracy than the \ac{ndi} controller. In fact, the \ac{mpc} approaches follow the set point nearly without delay. A reason for this behavior is the available preview information about the future set points. In this context, the error needs to occur first for the \ac{ndi} scheme to react. In contrast, \ac{mpc} can counteract a predicted error before it occurs. This becomes visible in the motor torque plots as well, where the curve shapes are comparable but have slight offset. Despite the generally better performance, \ac{mpc} shows a drift in the lateral position, and values above $2.5 \,\ur{mm}$ are not matched well. A possible explanation is the strongly simplified prediction model. 
		The control scheme with the best performance in this scenario is \ac{nmpc}, showing the potential capability of \ac{mpc} controllers for lateral railway running gear control.
		However, the accuracy comes at the cost of a high computational burden. The mean time to solve an optimization problem took $20.3 \,\ur{ms}$ in \ac{nmpc}. In contrast, \ac{ltvmpc} is slightly less accurate than \ac{nmpc}, but the computation time reduces with $5.6 \,\ur{ms}$ approximately to a quarter. This indicates that running gear control schemes can benefit from the conceptual advantages of \ac{mpc} even if computationally efficient \ac{mpc} variants are employed. 
		Similar results regarding control performance and computation time have been obtained in co-simulations at velocities from $40\,\ur{km/h}$ to $400\,\ur{km/h}$ based on the evaluation tracks $\cali{T}_1$ through $\cali{T}_5$.\\
		For comparison of the pure set point tracking performance, co-simulations based on evaluation track $\cali{T}_5$ and fixed sinusoidal set point trajectories for $y^{*}$ are conducted. As can be seen in \fig{figure_7}, the results resemble the previous qualitative observations regarding the lateral position \ac{rmse}. The most accurate lateral motion is achieved with \ac{nmpc}, followed by \ac{ltvmpc} and \ac{ndi}. Within each method, better results are obtained with lower velocities. The velocity dependence of the absolute lateral position error is highest with the \ac{ndi} method and is mitigated by the \ac{mpc}-based controllers.
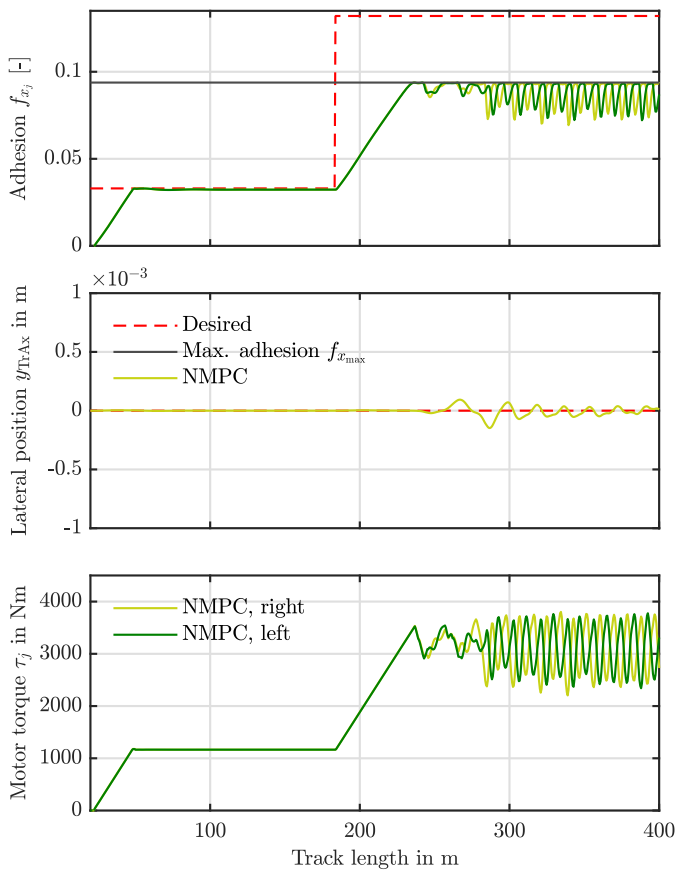
		\newfig[t]{width=0.49\textwidth}{figure_8}{Co-simulation result, evaluation track $\cali{T}_5$, $400 \ur{km/h}$, critical adhesion conditions. The longitudinal adhesion controller is able to conduct regular control and maximum-seeking control within one control law. No knowledge of the adhesion-slip characteristic is needed. Even with poor adhesion conditions longitudinal adhesion control and lateral position control is performed safely.}
		\subsection{Evaluation of Longitudinal and Integrated Control}\label{sec:evaluation_longitudinal}
		The proposed integrated control scheme employs a longitudinal subsystem controller based on a maximum-seeking control method, which has proven its superiority to state-of-the-art approaches in previous publications \cite{Schwarz.2021}. In \secref{sec:controller-synthesis}, the maximum-seeking control law is designed to handle both, regular operation (\ie $f_{x_j}^{*}<f_{x_{\ur{max}}}$) and critical operation situations (\ie $f_{x_j}^{*}>f_{x_{\ur{max}}}$). In \fig{figure_8}, the working principle is illustrated qualitatively based on evaluation track $\cali{T}_5$.\\
		First, a small traction force $F_x^{*}$ set point is provided such that $f_x^{*} < f_{x_{\ur{max}}}$. In this regular operation case, the motor torque converges to $f_x^{*}$, as proved in \secref{sec:adhesion-control}. From $130 \,\ur{m}$, the set point adhesion exceeds the maximum possible adhesion $f_{x_{\ur{max}}}$. In this situation, the controller should steer the adhesion to the maximum possible value while avoiding excessive slip and thus unstable behavior. As can be seen in the top plot of \fig{figure_8} from $240\,\ur{m}$, the maximum adhesion is indeed approached. As soon as one wheel-rail contact enters the macro slip regime, the control input $u^{\ur{d}}$ is decreased rapidly, which drives the system back to the stable micro slip region. The maximum adhesion is then approached again. 
		It is worth mentioning that the devised adhesion controller operates robustly without knowledge of the adhesion-slip characteristic and relies solely on basic assumptions. Regarding the lateral direction, the control task becomes more difficult due to the poor adhesion conditions. Nonetheless, the results in the middle plot demonstrate that \ac{nmpc} ensures a stable oscillation around the set point.\\
		To quantitatively show the effectiveness of the integrated control scheme, a simulation employing evaluation track $\cali{T}_2$ and a dynamic set point trajectory $y^*$ is conducted (see \fig{figure_9}). A high braking force set point $F_x^*$ is demanded and causes actuator saturation of the wheel drives. Thus, longitudinal and lateral control need to be performed with limited available control torque $\tau_j$ while maximization of braking force and minimization of the lateral displacement error are conflicting control objectives. 
		In this context, the performance criteria are the braking distance until standstill (longitudinal control) and the \ac{rmse} between desired lateral displacement $y^{*}$ and actual lateral displacement $\ytrax$ of the running gear with respect to the track center line (lateral control).\\ 
		To show the conceptual advantage of \ac{mpc}-based integrated control, we employ \ac{nmpc}, as the most accurate \ac{mpc}-based scheme, with two different parameter sets, exploiting the flexibility of \ac{mpc} to vary the control parameters $\bi{Q}$ and $R$ at custom, smoothly, and with interpretable impact on desired system variables.
		For performance comparison, no state-of-the-art controller for integrated longitudinal and lateral control of railway running gears with \ac{irw} is available. As an alternative, the \ac{ndi}-based lateral controller is combined with the longitudinal controller designed in \secref{sec:adhesion-control}.\\ 
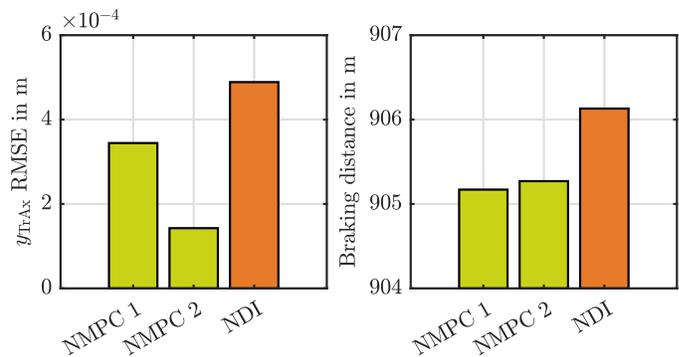
		\newfig[t]{width=0.49\textwidth}{figure_9}{Co-simulation result for \ac{nmpc}-based and \ac{ndi}-based integrated controllers with two parameter sets, $1000 \,\ur{m}$ of evaluation track $\cali{T}_{2}$, $160 \,\ur{km/h}$, good adhesion conditions, high braking demand and actuator saturation, sine sweep as $y^{*}$ set point. The results indicate that the proposed integrated controller allows for shifting the control emphasis smoothly between lateral and longitudinal control objective by adjusting control parameters. Both parametrizations of the integrated controller outperform the state-of-the-art \ac{ndi}-based scheme.}
		As can be seen in \fig{figure_9}, a trade-off between lateral accuracy and braking distance is possible employing the adjusted controller parameters. In this representative example, the only parameter changed is the input cost parameter $R$. 
		The key property, why \ac{mpc}-based lateral control is especially suited to be employed in the integrated controller, is the possibility to penalize deviations of states and inputs very specifically. Thus, the accuracy requirements and the input utilization in lateral \ac{mpc} can be varied smoothly and intuitively compared to other lateral control approaches (\eg \ac{ndi}). Apart from this methodological advantage, we see that both \ac{nmpc}-based controllers outperform the \ac{ndi} scheme in terms of lateral position accuracy and braking distance. 
		
		\section{Discussion}\label{sec:discussion}
		In this paper, \ac{mpc} for integrated control of railway running gears with \ac{irw} is explored for the first time. The results in \secref{sec:experimental-results} show the conceptual advantages of integrated control by means of \ac{mpc} in extensive \ac{mbs}. However, certain aspects require special attention.\\
		First, the employed prediction model relies on certain assumptions, such as conic wheels and ideal rolling. In our extensive \ac{mbs} evaluations, these simplifications were acceptable. In other contexts, conditions may differ, and model adjustments might be necessary.\\
		Second, the functionality of the adhesion-based controller is dependent on a set of basic assumptions. These assumptions may not be fulfilled in every single situation, which needs to be investigated in the future. Furthermore, both lateral and longitudinal subsystem controllers come with a number of control design parameters, which can partly influence each other so that tuning may be difficult.\\
		Last, the chosen evaluation methodology is primarily control-oriented in accordance with the main objectives to develop an integrated controller and to explore \ac{mpc} for railway running gear control. To enable practical application of the presented control methods, future research may build on the obtained results, focusing on practice-oriented evaluation scenarios and real-world experiments.\\
		Thus, the presented research findings introduce \ac{mpc}-based integrated control of railway running gears with \ac{irw} with focus on its methodological advantages, providing a solid basis for applied future research.

		\section{Conclusion and Outlook}\label{sec:conclusion}
		In the current work, we propose an integrated controller for high-speed railway running gears with driven \ac{irw}. Two subsystem controllers are employed and joined with a rule-based integration approach. Beyond the current state of research, the use of \ac{mpc} is explored for lateral railway running gear control. Key methodological advantages compared to existing approaches are the possibility to include preview information inherently in the prediction horizon and to smoothly shift the control objective of the integrated controller by adjusting \ac{mpc} parameters. For longitudinal control, a novel adhesion-based controller is developed. It can perform both, regular and maximum-seeking adhesion control without knowledge of the adhesion-slip characteristic or a specific contact model. 
		Thus, a holistic control scheme for handling of multiple regular and critical operation scenarios is contributed.\\
		Simulation results with a detailed \ac{mbs} train model show the effectiveness of the integrated controller. In particular, the presented control scheme outperforms a state-of-the-art \ac{ndi} controller. The results indicate that \ac{mpc} is a promising approach for integrated control of railway running gears with driven \ac{irw}. Specifically, the ability of \ac{mpc} to track highly dynamic set points suggests that it can counteract known lateral track irregularities actively in future implementations. This can help to increase riding safety and comfort by improving lateral control stability and avoiding wheel flange contact.\\
		A common drawback of \ac{mpc} is the computational burden. A \ac{ltvmpc} controller is devised to address the issue. An approximate optimization time reduction to a quarter of the initial value is achieved, indicating the feasibility of \ac{mpc} concepts for railway running gear control from a computational standpoint. Future investigations should focus further on real-time capable \ac{mpc} designs as well as practice-oriented implementations. Another interesting option for future research is the design of a centralized \ac{mpc} for joint handling of lateral and longitudinal control. 
		
		\section*{Acknowledgments}
		Jan-Hendrik Ewering would like to thank Tobias Posielek for active discussions and bringing in his knowledge about control systems theory and scientific writing. In addition, Jan-Hendrik Ewering would like to thank the research group leaders Andreas Heckmann and Zygimantas Ziaukas who enabled the collaboration between the Institute of Mechatronic Systems and the Institute of System Dynamics and Control.

		\bibliographystyle{IEEEtran}
		\bibliography{IEEEabrv,references}
%		\balance
		
	\end{document}